\documentclass[11pt]{article}%
\usepackage{hyperref}
\usepackage{latexsym}
\usepackage{amssymb,amsfonts,amsmath}
\usepackage{graphicx}
\usepackage{indentfirst}
\usepackage{amsmath}
\usepackage{amsfonts}
\usepackage{amssymb}%
\usepackage{comment}
\ifx\pdfoutput\undefined
\else
\fi
\hypersetup{colorlinks=false,bookmarksopen,bookmarksnumbered,citecolor=blue,
pdfstartview=FitH}

\parskip=\medskipamount

\def\a{\alpha}

\newcommand{\be}{\begin{equation}}
\newcommand{\ee}{\end{equation}}
\newcommand{\bea}{\begin{eqnarray}}
\newcommand{\eea}{\end{eqnarray}}

\newcommand{\ba}{\begin{array}}
\newcommand{\ea}{\end{array}}

\def\double #1{#1{\hbox{\kern-2pt $#1$}}}

\newcommand{\bsubeq}{\begin{subequations}}
\newcommand{\esubeq}{\end{subequations}}

\newcommand{\virgolette}{``}

\DeclareMathAlphabet{\mathpzc}{OT1}{pzc}{m}{it}

\begin{document}

\begin{titlepage}
\begin{flushright}
\par\end{flushright}
\vskip 1.5cm
\begin{center}
\textbf{\huge \bf Self-Dual Forms in Supergeometry I:}
\\
\vskip .2cm
\textbf{\Large \bf The Chiral Boson}

\vskip 1.5cm
\large {\bf C.~A.~Cremonini}$^{~a,b,}$\footnote{carlo.alberto.cremonini@gmail.com}, 
\large {\bf P.~A.~Grassi}$^{~c,d,e,}$\footnote{pietro.grassi@uniupo.it}, 
{\small
\vskip .5cm
\medskip
\centerline{$^{(a)}$ \it Dipartimento di Scienze e Alta Tecnologia (DiSAT),}
\centerline{\it Universit\`a degli Studi dell'Insubria, via Valleggio 11, 22100 Como, Italy}
\medskip
\centerline{$^{(b)}$ \it INFN, Sezione di Milano, via G.~Celoria 16, 20133 Milano, Italy} 
\medskip
\centerline{$^{(c)}$
\it Dipartimento di Scienze e Innovazione Tecnologica (DiSIT),} 
\centerline{\it Universit\`a del Piemonte Orientale, viale T.~Michel, 11, 15121 Alessandria, Italy}
\medskip
\centerline{$^{(d)}$
\it INFN, Sezione di Torino, via P.~Giuria 1, 10125 Torino, Italy}
\medskip
\centerline{$^{(e)}$
\it Arnold-Regge Center, via P.~Giuria 1,  10125 Torino, Italy}
}
\end{center}
\vskip  0.1cm
\begin{abstract}
Recent results of A. Sen on quantum field theory models with self-dual field strengths use 
string field theory as a starting point. In the present work, we show that combining string field 
theory and supergeometry we can provide a constructive method for all these models, for 
any superspace representation and for any given background. The analysis is based on 
the new concept of pseudoform, emerging in supergeometry, which opens a new page 
in quantum field theory and, in particular, in supergravity. The present work deals with an explicit example, the case of the chiral boson multiplet in $d=2$.

\end{abstract}
\vfill{}
\vspace{1.5cm}
\end{titlepage}
\newpage\setcounter{footnote}{0}
\tableofcontents

\section*{Introduction}\label{sec:intro}

\addcontentsline{toc}{section}{\nameref{sec:intro}}

Recently the work of A. Sen \cite{Sen:2015nph,Sen:2019qit} has brought back the attention to the 
long standing problem of constructing a consistent field theory with self-dual field strengths. There are several examples of 
models such as $N=2$, $d=10$ Type $IIB$ supergravity (where the 5-form field strength for the Ramond-Ramond (RR) fields has to be self-dual), $d=6$ tensor multiplet, 
self-dual Yang-Mills and the chiral boson. Several proposals have been considered, see for example the non-exhaustive list \cite{Floreanini:1987as,Henneaux:1988gg,Bastianelli:1989cu,Bastianelli:1989hi,McClain:1990sx,Wotzasek:1990zr,Devecchi:1996cp,Berkovits:1996nq,Berkovits:1996em,Pasti:1996vs,Pasti:1997gx,Schwarz:1997mc,DallAgata:1997gnw,DallAgata:1998ahf} and the more recent \cite{Lambert:2019diy,Lambert:2019khh,Buratti:2019guq,Mkrtchyan:2019opf,Andriolo:2020ykk,Bandos:2020hgy,1837177} (see in particular \cite{Townsend:2019koy}, which is mainly focused on the two-dimensional case) where the self-duality is a crucial requirement. 
Moreover, there 
are several {\it ad hoc} solutions which might solve the issue for special cases of interest (see e.g. \cite{AlvarezGaume:1983ig,Berkovits:1994wr,Berkovits:1996bf}), for example, one could violate manifest Lorenz invariance, or use an infinite number of auxiliary fields, or adopt non-polynomial Lagrangians. 
Here we consider a geometrical approach in order to deal with the supersymmetric version of the problem, which can be easily generalized to any background of supergravity.
A full-fledged action whose variational principle provides the correct covariant (finite number of) equations was missing until the proposal of A.Sen. 
The main problem is that, by a simple analysis of the equations of motion, one immediately sees that for each self-dual propagating degree of freedom, 
its anti-self-dual companion propagates as well, jeopardizing the counting of degrees of freedom. Sen's approach \cite{Sen:2015nph,Sen:2019qit} is different: 
it is based on the structure of superstring field theory action 
where the closed string sector (with Neveu-Schwarz-Neveu-Schwarz (NSNS) and RR fields) is described by two string fields, $\Phi$ and $\Psi$. The equations 
of motion fix them in terms of each other such that the self-dual part of the 5-form of RR couples to the rest of theory 
while the anti-self dual part decouples. The form of the action given in \cite{Sen:2015uaa} is
\begin{eqnarray}
\label{streA}
S_{_{SFT}} = \langle \Phi, Q \mathbb{Y} \Phi \rangle + \langle \Phi , Q \Psi \rangle + f(\Psi) \ , 
\end{eqnarray}
where $Q$ is the BRST operator, $\Phi$ and $\Psi$ are the string fields, $\mathbb{Y}$ is the Picture Changing Operator needed to provide a consistent kinetic term for the field $\psi$ and $f(\Psi)$ is a potential term for the string field $\Psi$. The field $\Psi$ is endowed with self-duality properties and therefore the coupling between $\Phi$ and $\Psi$ 
determines which part of the field $\Phi$ couples to the rest of the theory. In particular, as shown in  \cite{Sen:2015nph,Sen:2019qit}, if 
$\Psi$ represents a self-dual form, only the self-dual part of $\Phi$ couples to the rest of the theory and the anti-self dual part of it decouples from the rest, or viceversa. 

We will tackle the problem of self-dual forms within the geometrical framework of \emph{rheonomy} \cite{cube}; in particular this framework is efficient for (rigid) supersymmetric and supergravity models. The key point of this language is that it allows to construct Lagrangians in terms of differential forms, hence keeping manifest the symmetries of the theory. Rheonomy allows to study a large variety of models such as extended supergravity ($N=2,$ type $II B$ supergravity of $d=6$ supergravity), where one has eventually to face the problem of self-dual field strengths. Some of the models are discussed in the books \cite{cube}, others in some articles \cite{Castellani:1991jf,Castellani:1993ye}. However, despite the power of the framework, even the simple case of chiral boson and its superpartner was not properly addressed.\footnote{To our knowledge, the case of the chiral boson has not been analysed in the rheonomic framework, nevertheless in several works (\cite{cube,Castellani:1991jf,Castellani:1993ye}) the authors propose some solutions to deal with self-dual field strengths. Those solutions are not satisfactory since the rheonomic variation principle does not include self-dual conditions.}
This is due to the fact that the rheonomic Lagrangian has to be supplemented with a \emph{Picture Changing Operator} (PCO) in order to define a consistent action principle. The PCO considered in rheonomy is the component one, projecting the Lagrangian to its component version or analogously setting the odd variables $\theta$'s and their even differentials $d \theta$'s to zero. Being the geometric PCO a de Rham cohomology representative, this choice is not at all unique. The problem in this context is that once the Lagrangian is projected via \emph{any} PCO, the resulting equations of motion do not correspond to the single chiral boson as one would expect. We will show that this problem is related to the fact that the Maurer-Cartan (MC) equations (once the conventional constraint has been implemented) imply the equations of motion. Hence, the verification of the closure of the Lagrangian is \virgolette trivialised" in the sense that the MC equations put the Lagrangian on-shell. This is a long-standing problem in rheonomy and it is tighten to the existence of auxiliary fields.

Before the implementation of the SFT-inspired method for describing self-dual forms, we will recall the construction of the Hodge dual on supermanifolds discussed in \cite{CCGir, Castellani:2015ata}. This is built on the complete complex of forms, it is an involutive operation and it acts as
\begin{eqnarray}
\label{HoA}
\star: \Omega^{(p|q)} \left( \mathcal{SM}^{(2n|2m)} \right) \rightarrow  \Omega^{(2n-p|2m-q)} \left( \mathcal{SM}^{(2n|2m)} \right) \ ,
\end{eqnarray}
and therefore (anti) self-dual forms can be defined only if $p=n$ and $q=m$. Self-dual forms on supermanifolds  are then immediately found to live in the \emph{psuedoform complex} $\Omega^{(n|m)}$. They can be lifted to integral forms by means of the PCO $\mathbb{Y}$, or be de-lifted to superforms by means of the PCO $Z$. 

With all these ingredients, we are able to provide the action on a supermanifold $\mathcal{SM}^{(2n|2m)}$. Given a $(n-1|0)$-form $\Phi$, we have
\begin{eqnarray}\label{streAA}
S = \int_{_{\mathcal{SM}^{(2n|2m)}}} \hspace{-1cm} \left( \mathcal{L}^{(2n|0)} (\Phi) \wedge \mathbb{Y}^{(0|2m)} + d\Phi^{(n-1|0)} \wedge Q^{(n|2m)} \right) \ ,
\end{eqnarray}
where $Q^{(n|2m)}$ is a \virgolette self-dual form" (actually, it is an integral form obtained from a self-dual form, as will be explained in the following). The first term reproduces the kinetic term for the field $\Phi^{(n-1|0)}$, while the second term reproduces Sen's coupling between $\Phi^{(n-1|0)}$ and the rest of the theory. The form of the action resembles, at the level of quantum field theory on a supermanifold, the closed string field theory action (\ref{streA}). In the particular case of the chiral boson we will set $n=1, m=1$.

In the first section, we will review the rheonomic Lagrangian for the chiral boson in $N=1$,$d=2$, and the related problems. In the second section, we will give a self-contained description of self dual (pseudo)forms on a supermanifold $\mathcal{SM}^{(2|2)}$. We will omit a description of form complexes on supermanifolds, for a complete discussion we refer the reader to \cite{Witten:2012bg,Catenacci:2018xsv,Catenacci:2019ksa,Cremonini:2019aao,Cremonini:2019xco,Cacciatori:2020hcm}. In the third section, we will present the the SFT-inspired action and the counting of degrees of freedom. We speculate how to rewrite the action in the pseudoform language, this will be object of future investigations.

\section{$N=1 , d=2$ Flat Supermanifolds}

We will begin by reviewing both the $N=1$,$d=2$ non-chiral model and the chiral one. We will briefly show how to construct the Lagrangian from the rheonomic prescriptions and the resulting (apparent, as we will discuss in full details) equations of motion.

\subsection{The Non-chiral Model}

Let us briefly explain the basic geometry. We consider two bosonic coordinates $(z, \bar z)$ (defined as complex coordinates) and two Grassmann odd coordinates $(\theta, \bar\theta)$, corresponding to the superspace $N = 1$, $d = 2$. We also introduce the differentials $(dz, d\bar z, d\theta, d\bar\theta)$ and the flat supervielbeine 
\begin{eqnarray}
\label{oneA}
V = dz + \theta d\theta\,, ~~~~~~
\bar V = d\bar z + \bar \theta d\bar\theta\,, ~~~~~~ \psi = d\theta\,,~~~~~
\bar\psi = d \bar \theta\,, 
\end{eqnarray}
which are supersymmetric invariant under $\delta \theta = \epsilon, \delta \bar{\theta} = \bar{\epsilon}$ and $\delta z = \epsilon \theta, \delta \bar{z} = \bar{\epsilon} \bar{\theta}$:
\begin{eqnarray*}
	\delta V &=& d \left( \delta z \right) + \delta \theta d \theta + \theta d \left( \delta \theta \right) = - \epsilon d \theta + \epsilon d \theta + \theta d \epsilon = 0 \ , \\
	\delta \bar{V} &=& d \left( \delta \bar{z} \right) + \delta \bar{\theta} d \bar{\theta} + \bar{\theta} d \left( \delta \bar{\theta} \right) = - \bar{\epsilon} d \bar{\theta} + \bar{\epsilon} d \bar{\theta} + \bar{\theta} d \bar{\epsilon} = 0 \ , \\
	\delta \psi &=& d \left( \delta \theta \right) = d \epsilon = 0 \ , \\
	\delta \bar{\psi} &=& d \left( \delta \bar{\theta} \right) = d \bar{\epsilon} = 0 \ , 
\end{eqnarray*}
where we have used the odd parity of $\epsilon$ and $\bar{\epsilon}$ and the fact that we are considering rigid supersymmetries $d \epsilon = 0 = d \bar{\epsilon}$.
The forms in \eqref{oneA} satisfy the Maurer-Cartan algebra 
\begin{eqnarray}
\label{oneAB}
d V = \psi\wedge \psi\,, ~~~~~~
d \bar V = \bar\psi\wedge \bar\psi\,, ~~~~~~ d\psi =0\,, ~~~~~~
d\bar\psi =0\,. 
\end{eqnarray}
The corresponding algebra of vector fields is given by
\begin{equation}\label{algebra2d}
	D = \partial_\theta -  \theta \partial_z \ , \ \bar D = \partial_{\bar\theta} - \bar\theta \bar\partial_{\bar{z}} \ , \ D^2 = - \partial_z \ , \ \bar D^2 = - \bar\partial_{\bar{z}} \ , \ \left[ D , \bar{D} \right] = D \bar D + \bar D D=0 \ .
\end{equation}
In order to avoid clumsy notations we will denote $\partial_z \equiv \partial$ and $\bar{\partial}_{\bar{z}} \equiv \bar{\partial}$.

Let us now start by describing the non-chiral multiplet. This is given by a superfield $\Phi$ with the decomposition 
\begin{eqnarray}
\label{oneB}
\Phi = \phi +  \phi_\theta \theta + \phi_{\bar\theta} \bar\theta + \phi_{\theta\bar\theta} \theta \bar\theta \,, ~~~~
W = D \Phi \,, ~~
\bar W = \bar D \Phi\,,  ~~
F =  \bar D D \Phi\,.
\end{eqnarray}
The component fields $\phi (z,\bar z),  \phi_\theta (z,\bar z) ,  \phi_{\bar\theta}(z,\bar z) $ and $ \phi_{\theta\bar\theta}(z,\bar z) $ are the spacetime fields; the first and last are even while the second and third are odd. On the other hand, $(\Phi, W, \bar W, F)$ are the superfields whose first components are the components fields enumerated before. $\Phi$ and $F$ are even while $W$ and $\bar W$ are odd superfields. 

If we write the differential of each superfield we find the following relations (we use $\displaystyle d = V \partial + \bar{V} \bar{\partial} + \psi D + \bar{\psi} \bar{D}$)
\begin{eqnarray}
\label{oneC}
d \Phi &=& V \partial \Phi + \bar V \bar\partial \Phi + \psi W + \bar\psi \bar W\,, \nonumber \\
d W &=& V \partial W + \bar V \bar\partial W - \psi \partial \Phi + \bar \psi F\,, \nonumber \\
d \bar W &=& V \partial \bar W + \bar V \bar\partial \bar W - \bar\psi \bar \partial \Phi - \psi F\,, \nonumber \\
d F &=& V \partial F + \bar V \bar\partial F + \psi \partial \bar W - \bar \psi \bar\partial W\,. 
\end{eqnarray}
The last field $F$ is the auxiliary field and therefore we expect its equation of motion to be purely algebraic and to set it to zero. Before we write the rheonomic Lagrangian for the multiplet, let us first write down the equations of motion. If we set $F =0$, we see from the last equation of \eqref{oneC} that 
\begin{eqnarray}
\label{oneD}
\partial \bar W =0\,, ~~~~~
\bar \partial W =0\,. 
\end{eqnarray}
These equations imply that the superfield $W$ is holomorphic $W = W(z)$ and its conjugated $\bar W$ is anti-holomorphic. If we now substitute these constraints in (\ref{oneC}) we obtain
\begin{eqnarray}
\label{oneE}
d \Phi &=& V \partial \Phi + \bar V \bar\partial \Phi + \psi W + \bar\psi \bar W\,, \nonumber \\
d W &=& V \partial W  - \psi \partial \Phi \,, \nonumber \\
d \bar W &=& \bar V \bar\partial \bar W - \bar\psi \bar\partial \Phi \,.
\end{eqnarray}
We can now impose the nilpotence condition for the exterior derivative (corresponding to the Bianchi identities) $d^2 =0$ leading to $\partial \bar \partial \Phi =0$. Hence, the full set of equaions of motion reads
\begin{eqnarray}
\label{oneF}
\partial \bar \partial \Phi =0 \,, ~~~~~~~
\partial \bar W =0\,, ~~~~~~~
\bar \partial W =0\,, ~~~~~~~
F =0\,. 
\end{eqnarray}
which are the free equations of $N=1,d=2$ supermultiplet. The Klein-Gordon equation in $d=2$ implies that the superfield $\Phi$ is decomposed into holomorphic and anti-holomorphic parts:
\begin{equation}
	\partial \bar{\partial} \Phi = 0 \ \implies \  \Phi = \Phi_z + \Phi_{\bar z}\ ,
\end{equation}
and therefore we get the on-shell matching of the degrees of freedom. In particular, we can write the on-shell holomorphic and anti-holomorphic superfields as
\begin{eqnarray}
\label{oneG}
\Phi_z = \phi(z) + \phi_{\theta}(z) \theta\,, ~~~~~
\Phi_{\bar z} = \phi(\bar z) +  \phi_{\bar\theta}(\bar z)\bar\theta\,, ~~~~~
\end{eqnarray}
factorizing into left- and right-movers. \\

Let us now focus in writing the action. In order to write an action in the first order formalism, we can introduce two additional auxiliary superfields $\xi$ and $\bar\xi$, to be used as Lagrange multipliers. Then, the action reads
\begin{eqnarray}
\label{oneH}
{\cal L}^{(2|0)} &=& (\xi V + \bar \xi \bar V) \wedge ( d \Phi - \psi W - \bar \psi \bar W) + 
\left(\xi \bar \xi + \frac{F^2}{2}\right)V\wedge \bar V + \nonumber \\
&+& 
W dW \wedge V - \bar W d \bar W \wedge \bar V - d\Phi \wedge (W \psi - \bar W \bar \psi) 
- W \bar W \, \psi \wedge \bar \psi \ .
\end{eqnarray}
The corresponding equations of motion can be easily calculated, so that they read
\begin{eqnarray}
\label{oneHA}
&&V \wedge (d\Phi - \psi W - \bar\psi \bar W) + \bar \xi V\wedge \bar V =0 \ , \nonumber \\
&&\bar V \wedge (d\Phi - \psi W - \bar\psi \bar W) + \xi V\wedge \bar V =0 \ , \nonumber \\
&&(\xi V + \bar \xi \bar V) \psi + 2 d W \wedge V - W \psi\wedge \psi + d\Phi \wedge \psi - \bar W \psi \wedge \bar\psi=0 \ , \nonumber \\
&&(\xi V + \bar \xi \bar V) \bar\psi - 
2 d \bar W \wedge V + \bar W \bar\psi\wedge \bar\psi - d\Phi \wedge \bar\psi + W \psi \wedge \bar\psi=0 \ , \nonumber \\
&&d ( \xi V + \bar \xi \bar V) + d W \psi - d\bar W\bar \psi =0 \ , \nonumber \\
&& F = 0 \ . 
\end{eqnarray}
It is an easy task to see that they imply the on-shell differentials (\ref{oneE}), the equations of motion (\ref{oneF}) and the relations for the new auxiliary fields in terms of $\Phi$
\begin{eqnarray}
\label{oneHB}
\xi  = \partial \Phi \ , ~~~~~
\bar \xi = - \bar\partial \Phi \ . 
\end{eqnarray}
The rheonomic action is a $(2|0)$ superform. It can be verified that it is closed, only by using the algebraic equations of motion for $\xi$ and $\bar \xi$ (\ref{oneHB}) and the curvature parametrisation $d\Phi, dW, d\bar W$ and $d F$ given in (\ref{oneC}). Note that those equations are off-shell parametrizations of the curvatures and therefore they do not need the equations of motion stemming out of (\ref{oneH}) (this a priori \virgolette trivial" consideration will be exactly the origin of the problem arising in the chiral case).

To move from the rheonomic Lagrangian to the component one we have to use use the component PCO ${\mathbb Y}^{(0|2)} = \theta \bar\theta \delta(\psi) \delta(\bar\psi)$ (we refer the reader to the Appendix for an introduction to integral forms, PCO's and the notations we use in the following sections):
\begin{eqnarray}
\label{oneI}
S &=& \int_{{\cal SM}} {\cal L}^{(2|0)}\wedge {\mathbb Y}^{(0|2)} = \nonumber \\
&=&  
\int_{z,\bar{z},dz,d\bar{z}} \left[ 
(\xi_0 dz + \bar\xi_0 d\bar z)\wedge d \phi + \left(\xi_0 \bar \xi_0 + \frac12 {\phi_{\theta\bar\theta}^2} \right) dz\wedge d\bar z + 
\phi_\theta d \phi_\theta \wedge dz + \phi_{\bar\theta} d \phi_{\bar\theta} \wedge d\bar z \right] = \nonumber \\
&=& \int_{z,\bar{z}} \left[ 
\xi_0 \partial\phi - \bar\xi_0 \bar{\partial} \phi + \xi_0 \bar \xi_0 + \frac12 {\phi_{\theta\bar\theta}^2} - \phi_\theta \bar{\partial} \phi_\theta + \phi_{\bar\theta} \partial \phi_{\bar\theta} \right] \ ,
\end{eqnarray}
where $\xi_0$ and $\bar \xi_0$ are the first components of the superfields $\xi$ and $\bar \xi$. Due to the choice of the component PCO, we have used the Dirac delta's to project out all pieces depending on the gravitinos $\psi, \bar{\psi}$ and the term $\theta \bar\theta$ to project out the dependence on Grassmann coordinates. If we eliminate $\xi_0$, $\bar \xi_0$ and $\phi_{\theta \bar{\theta}}$ via their algebraic equation of motion, we obtain the usual action of $d=2$ free sigma model \cite{Catenacci:2018jjj}:
\begin{equation}\label{oneIA}
	\int_{z,\bar{z}} \left[ \partial \phi \bar{\partial} \phi - \phi_\theta \bar{\partial} \phi_\theta + \phi_{\bar\theta} \partial \phi_{\bar\theta} \right] \ .
\end{equation} 

In order to consider the superspace action, we have to consider the supersymmetric PCO: 
\begin{eqnarray}
\label{oneL}
{\mathbb Y}^{(0|2)} = V \iota \delta(\psi) \wedge \bar V \bar \iota \delta(\bar\psi) \ . 
\end{eqnarray}
By substituting this PCO in the action we obtain
\begin{eqnarray}
\label{oneM}
S &=& \int_{\cal SM} \left[ W \bar W \psi \wedge \bar \psi - d\Phi \wedge (W \psi + \bar W \bar \psi)  \right] \wedge
V \iota \delta(\psi) \wedge \bar V \bar \iota \delta(\bar\psi) \nonumber \\
&=& - \int_{\cal SM} \Big( W \bar W - [\bar{D} \Phi W + D \Phi \bar W]\Big) V\wedge \bar V \wedge \delta(\psi) \delta(\bar \psi) = \nonumber \\
&=& \int_{z,\bar{z},\theta,\bar{\theta}} \Big[ - W \bar W + \bar{D} \Phi W + D \Phi \bar W \Big] \ .
\end{eqnarray}
If we eliminate the superfields $W$ and $\bar W$ via their algebraic equations of motion, we obtain the usual free $d=2$ superspace action in flat background:
\begin{equation}\label{oneN}
	\int_{z,\bar{z},\theta,\bar{\theta}} D \Phi \bar{D} \Phi \ .
\end{equation}
It is easy to verify that \eqref{oneN} leads to \eqref{oneIA} (after the Berezin integration). We want to stress the fact that we already knew a priori that the two actions describe the same field content, since the Lagrangian is closed, hence there is no dependence on the choice of the PCO (again, see the Appendix for further comments).

\subsection{The Chiral Model}

Let us now move to the chiral model. In this case we require that $\bar D \Phi =0$, which implies also that $\bar \partial \Phi =0$, due to the algebraic relations \eqref{algebra2d}. In this case, the rheonomic equations \eqref{oneC} become 
\begin{eqnarray}
\label{oneO}
d \Phi = V \partial \Phi + \psi W\,, ~~~~~
d W = V \partial W - \psi \partial \Phi\,. 
\end{eqnarray}
There is no room for the auxiliary field $F$ and the field content is represented by one scalar field $\phi$ and one fermion $w$ only (the first components of the superfields $\Phi$ and $W$ respectively). The chirality implies that $\Phi$ is also holomorphic (and that in turn implies that it is on-shell) and this matches with $w$ which is holomorphic on-shell. The free equations of motion are 
\begin{eqnarray}
\label{oneP}
\bar \partial \Phi =0\,, ~~~~~~~
\bar \partial W =0\,.  
\end{eqnarray}
Yet at this point, it appears clear that writing an action encoding both equations (\ref{oneO}) and (\ref{oneP}) might be difficult. The key point is that we cannot use (\ref{oneO}) directly in the action since they imply on-shell conditions. We will write down an action according to the rheonomic prescriptions, but we will then shown that it is not consistent (in the sense that it does not describe the desired chiral boson). 

Let us consider the rheonomic Lagrangian (the superspace is still $N=1$ and therefore, we still have the second coordinate $\bar\theta$)
\begin{eqnarray}
\label{oneQ}
{\cal L}^{(2|0)}_c = \Big[(\xi V + \bar\xi \bar V) \wedge (d\Phi - \psi W) + \xi \bar \xi V\wedge \bar V + 
W d W \wedge V -  d\Phi \wedge \psi W \Big] \ ,
\end{eqnarray}
obtained from \eqref{oneH} by setting to zero $\bar W$ and $F$. Here a problem arises: if we want to check whether the Lagrangian is closed, we have to use the MC \eqref{oneO}, but these equations imply the equations of motion. Therefore ${\mathcal L}^{(2|0)}_c(\Phi, W, V, \psi)$ is \virgolette trivially closed", in the sense that when we verify its closure we are immediately on-shell. This is a priori a non-trivial problem, since we are not ensured that any modification of the PCO will leave the theory unchanged.

In order to show the emergence of these problems from a different point of view, let us consider the equations of motion \emph{without} the PCO, i.e., the equations obtained from \eqref{oneQ} assuming the independence on the PCO (in other words, the closure of the Lagrangian). The equations of motion read
\begin{eqnarray}
\label{oneR}
V \wedge (d \Phi - \psi W) + \bar \xi V\wedge \bar V =0\,, \nonumber \\
\bar V \wedge (d \Phi - \psi W) + \xi V\wedge \bar V =0\,, \nonumber \\
d ( \xi V + \bar\xi \bar V) + d W \wedge \psi =0\,, \nonumber \\
\psi (\xi V + \bar \xi \bar V) + 2 d W\wedge V + d\Phi\wedge \psi =0\,. 
\end{eqnarray}

From the first equation, we have 
\begin{eqnarray}
\label{priA}
 W = D\Phi\,, ~~~~~
 \bar \xi = - \bar\partial \Phi\,, ~~~~~
 \bar D \Phi =0\,;
\end{eqnarray}
from the second equation, we get
\begin{eqnarray}
\label{priAA}
\xi = \partial \Phi\,; ~~~~~
\end{eqnarray}
from the third equation, we get
\begin{eqnarray}
\label{priAB}
&&\bar\partial \xi - \partial \bar \xi =0\,, ~~~~~
D \xi + \partial W =0\,, ~~~~~
\bar D \xi=0\,, ~~~~
D \bar \xi + \bar \partial W =0\,, ~~~~\nonumber \\
&&\xi + D W =0\,, ~~~~~
\bar D W =0\,, ~~~
\bar\xi =0\,. 
\end{eqnarray}
Combining all these equations, we obtain
\begin{eqnarray}
\label{priC}
&&W = D\Phi\,, ~~~~~ \xi = \partial \Phi\,, ~~~~ \bar \xi =0\,, ~~~~~ \bar D\Phi =0\,, \nonumber \\
&&\bar \partial \Phi=0\,, ~~~~~\bar D W =0\,, ~~~~~ \bar D \xi =0\,, \nonumber \\
&& \bar\partial W =0\,, ~~~~~\bar\partial \xi =0\,.  
\end{eqnarray}
From this set of equations we see that $W$ and $\xi$ are defined in terms of the derivatives of $\Phi$, and $\Phi$ is chiral $\bar D\phi=0$. Clearly, it follows the matching of bosonic and fermionic degrees of freedom.

The action on the whole supermanifold, corresponding to the previous Lagrangian reads
\begin{eqnarray}
\label{rheoB}
S = \int_{_{\mathcal{SM}}}{\mathcal L}^{(2|0)}_c(\Phi, W, V, \psi) \wedge \mathbb{Y}^{(0|2)} \ .
\end{eqnarray}
As we stressed many times, if the action were closed, we could change the PCO by exact terms without changing the field content, but since ${\mathcal L}^{(2|0)}_c(\Phi, W, V, \psi)$ is not closed (or worse, its MC imply the equations of motion), we cannot freely make such a choice. In the following paragraphs we will study explicitly the cases corresponding to some choices of the PCO. We will show that the \emph{correct} equations of motion coincide with a subset of \eqref{priC} and actually do not represent a single chiral boson (and its supersymmetric partner).

\subsubsection{The Component PCO}

Let us start by considering the component PCO ${\mathbb Y}^{(0|2)} = \theta \bar\theta \delta(\psi) \delta(\bar\psi)$. It projects the Lagrangian on a purely bosonic manifold and from \eqref{rheoB} we obtain
\begin{equation}
\label{oneRA}
S = \int_{\cal SM} {\cal L}^{(2|0)}_c \wedge \theta \bar\theta \delta(\psi) \delta(\bar\psi) = \int_{z,\bar{z}} \Big( \xi_0 \bar{\partial} \phi - \bar{\xi}_0 \partial \phi + \xi_0 \bar{\xi}_0 - w \bar{\partial} w \Big) \ .
\end{equation}
The action does not describe a chiral boson anymore: if we use the algebraic equations for $\xi_0$ and $\bar{\xi}_0$ we obtain the following equations of motion
\begin{equation}
	\partial \bar{\partial} \phi = 0 \ , \ \bar{\partial} w = 0 \ .
\end{equation}
In particular, we have a propagating holomorphic fermion, while both the holomorphic and the anti-holomorphic parts of the boson propagate.

\subsubsection{The Half-Supersymmetric PCO}

Let us consider another PCO of the form 
\begin{eqnarray}
\label{oneS}
{\mathbb Y}^{(2|0)} = V \iota \delta(\psi) \wedge \bar\theta \delta(\bar \psi) \ ,
\end{eqnarray}
which is manifestly supersymmetric on the holomorphic part only. It is easy to check that 
indeed is closed and not exact (see the Appendix for explicit calculations). Inserting it into the action we get 
\begin{equation}
\label{oneT}
S = \int_{\cal SM} {\cal L}^{(2|0)}_c \wedge V \iota \delta(\psi) \bar\theta \delta(\bar \psi) = \int_{\cal SM}  \Big[ \bar \xi (D \Phi + W) - \bar\partial \Phi W \Big] \bar\theta V\wedge \bar V \delta(\psi) \delta(\bar \psi) = $$ $$ = \int_{z,\bar{z},\theta} \Big[ \bar \xi (D \Phi + W) - \bar\partial \Phi W \Big] = \int_{z,\bar{z}} \Big[ D \bar \xi (D \Phi + W) + \bar\xi ( \partial \Phi + D W) + \partial \Phi \bar \partial \Phi - W \bar \partial W \Big]_{\theta=\bar \theta =0}\,. 
\end{equation}
This action leads to four equations of motion, two of which are algebraic. The resulting two equations of motion read
\begin{equation}\label{oneTA}
	\partial \bar{\partial} \phi = 0 \ , \ \bar{\partial} w = 0 \ ,
\end{equation}
exactly as in the previous case. The same field content would have been obtained by considering the conjugate PCO
\begin{eqnarray}
{\mathbb Y}^{(2|0)} = \theta \delta(\psi) \wedge \bar{V} \bar{\iota} \delta(\bar \psi) \ .
\end{eqnarray}

We can also consider a composite PCO as
\begin{eqnarray}
\label{oneUA}
{\mathbb Y}^{(2|0)} = V \iota \delta(\psi) \wedge \bar\theta \delta(\bar \psi) + 
\theta \delta(\psi) \wedge  \bar V \bar\iota \delta(\bar\psi)  
\end{eqnarray}
where the relative coefficient is fixed by requiring the reality of the PCO. This PCO leads to
\begin{eqnarray}
\label{oneW}
S = \int_{\cal SM}  \left[ \left( \bar \xi (D \Phi + W) - \bar\partial \Phi W \right) \bar\theta V\wedge \bar V \delta(\psi) \delta(\bar \psi) + \left( \xi \bar{D} \Phi - W \bar{D} W \right) \theta V \bar{V} \delta(\psi) \delta(\bar \psi) \right] \ ,
\end{eqnarray}
which can be seen to lead exactly to \eqref{oneTA}.

\subsubsection{The Supersymmetric PCO}

Finally, we can also use the supersymmetric PCO 
\begin{eqnarray}
\label{oneU}
{\mathbb Y}^{(0|2)} = V \iota \delta(\psi) \wedge \bar V \bar\iota \delta(\bar \psi) \,, 
\end{eqnarray}
we have 
\begin{equation}
\label{oneZ}
S = \int_{\cal SM} {\cal L}^{(2|0)}_c \wedge  V \iota \delta(\psi) \wedge \bar V \bar\iota \delta(\bar \psi) = \int_{z,\bar{z},\theta, \bar{\theta}} \left[ - \bar D\Phi W \right] \ ,
\end{equation} 
whose equations of motion are 
\begin{eqnarray}
\label{oneZA}
\bar D \Phi =0\,, ~~~~~~
\bar D W =0\,,
\end{eqnarray}
which imply that $\bar \partial \Phi=0$ and $\bar\partial W=0$ which are the equations of motion for two chiral superfields. Each equation of \eqref{oneZA} describes a chiral boson and a chiral fermion, hence leading to two chiral bosons and their corresponding superpartners. This example not only shows that again this PCO is not suitable for describing a single chiral boson (and its superpartner), but also that in this case the field content is represented by two holomorphic bosons and two holomorphic fermions, in contrast to the field content we found for the component or half-supersymmetric PCO's. Actually, this confirms that the action is not closed and that the PCO's always project a non-trivial part of the \virgolette would-be" equations of motion \eqref{priC}.

One interesting aspect that we have to consider is the following: using the supersymmetric PCO (\ref{oneU}) we have derived a superspace action (\ref{oneZ}) which is manifestly supersymmetric, being written in terms of superfields integrated on the superspace. Since the propagating degrees of freedom do not coincide with those of the component action, it is interesting to see which component action we obtain by performing the Berezin integration of the superspace one. Given the decompositions of the superfields $\Phi$ and $W$ 
\begin{eqnarray}
\label{decoAA}
\Phi = \phi + \phi_{\theta} \theta + \phi_{\bar{\theta}} \bar\theta + \phi_{\theta \bar{\theta}} \theta \bar\theta\,, ~~~~~~~~~
W = w + w_{\theta} \theta + w_{\bar{\theta}} \bar\theta + w_{\theta \bar{\theta}} \theta \bar\theta\,, ~~~~~~~~~
\end{eqnarray}
where the Grassmannian nature of the fields is understood (recall that the superfield $\Phi$ is even and the superfield $W$ is odd), we can compute the component action as
\begin{eqnarray}
\label{selfdecoC}
S = \int_{z,\bar{z}} \left( - \phi_{\theta} w_{\theta \bar{\theta}} + \phi_{\theta \bar{\theta}} w_{\bar{\theta}} - \bar{\partial} \phi w_{\theta} - \bar{\partial} \phi_{\theta} w \right) \ ,  
\end{eqnarray}
where the Berezin integral has already been performed. After removing the algebraic equations, we are left with
\begin{eqnarray}
\label{decoDA}
S = - \int_{z,\bar{z}} \left( \bar{\partial} \phi w_{\theta} + \bar{\partial} \phi_{\theta} w \right)\,,   
\end{eqnarray}
which is a Lagrangian for a $bc-\beta\gamma$ system used in the BRST quantization of NRS superstring. It is supersymmetric and it propagates two chiral bosons $\phi$ and $w_\theta$ and two chiral fermions $w$ and $\phi_\theta$. \\

The goal of these subsections was to show that the rheonomic formulation of the chiral boson is affected by some non-trivial and intrinsic problems. The point is that the \virgolette naive" equations of motion \eqref{oneR} are incorrect. The correct equations of motion are
\begin{eqnarray}
\label{correcteom}
&&\Big( V \wedge (d \Phi - \psi W) + \bar \xi V\wedge \bar V  \Big)\wedge \mathbb{Y}^{(0|2)} =0 \,, \nonumber \\
&&\Big( \bar V \wedge (d \Phi - \psi W) + \xi V\wedge \bar V \Big)\wedge \mathbb{Y}^{(0|2)} =0 \,, \nonumber \\
&&\Big( d ( \xi V + \bar\xi \bar V) + d W \wedge \psi  \Big)\wedge \mathbb{Y}^{(0|2)}  =0\,,  \nonumber \\
&&\Big( \psi (\xi V + \bar \xi \bar V) + 2 d W\wedge V + d\Phi\wedge \psi \Big)\wedge \mathbb{Y}^{(0|2)} =0\,,
\end{eqnarray}
that is, we have to consider the projection with the PCO as well. This means that not all the equations of motion will survive the projection. Despite the huge redundancy of \eqref{oneR}, we have seen that any choice of PCO leave untouched an \virgolette insufficient" number of equations of motion, whose solution do not coincide with the solution of \eqref{oneR}. When considering theories with closed Lagrangians, we have that \emph{any} choice of the PCO selects a subset of the rheonomic equations of motion whose solution coincides with the one of the full rheonomic set.\\

Hence the necessity of finding an alternative, consistent solution for treating the chiral boson Lagrangian (and in general any Lagrangian containing self-dual forms). In the following sections we will implement the strategy pioneered by A.Sen in \cite{Sen:2015nph,Sen:2019qit} to our supergeometric context, showing that it actually solves the problem of self-duality in field theory.

\section{Self-Dual Forms in Supergeometry}

Before we describe the implementation of the SFT-inspired method, let us briefly recall some basic definitions and notations about self-duality in supergeometry. We will focus mainly on the case of our interest $\mathcal{SM}^{(2|2)}$, while these results are ready for higher dimensional cases. We will avoid general treatment of Hodge theory on supermanifolds (which can be found in \cite{Castellani:2015ata}), we will give basic definitions and results for the $\star$ operator in order to make the paper self-contained.

In this section we will focus on the \virgolette flat operator", which is defined as
\begin{eqnarray}\label{ADHAA}
	\star : \Omega^{(p|q)} \left( \mathcal{SM}^{(n|m)} \right) &\rightarrow& \Omega^{(n-p|m-q)} \left( \mathcal{SM}^{(n|m)} \right) \nonumber \\
	\omega^{(p|q)} &\rightarrow& \star \omega^{(p|q)} \ , \ \text{such that} \  \omega^{(p|q)} \wedge \star \omega^{(p|q)} = \mathpzc{B}er^{(n|m)} \ ,
\end{eqnarray}
where $\mathpzc{B}er^{(n|m)}$ is the volume form of the supermanifold. Notice that actually we are considering the action of the $\star$ operator leaving the components of the forms unchanged. When writing $\omega^{(p|q)} \wedge \star \omega^{(p|q)} = \mathpzc{B}er^{(n|m)}$ we are thus assuming the components to be constants (or even to be 1). If we restrict to the case of $n=2=m$, we have
\begin{eqnarray}\label{ADHA}
	\star : \Omega^{(p|q)} \left( \mathcal{SM}^{(2|2)} \right) &\rightarrow& \Omega^{(n-p|m-q)} \left( \mathcal{SM}^{(2|2)} \right) \nonumber \\
	\omega^{(p|q)} &\rightarrow& \star \omega^{(p|q)} \ , \ \text{such that} \  \omega^{(p|q)} \wedge \star \omega^{(p|q)} = \mathpzc{B}er^{(2|2)} = V \bar{V} \delta \left( \psi \right) \delta \left( \bar{\psi} \right) \ .
\end{eqnarray}
From the definition \eqref{ADHAA} it follows the involution property of the $\star$ operator on a supermanifold $\mathcal{SM}^{(n|m)}$:
\begin{equation}\label{ADHC}
	\mathpzc{B}er^{(n|m)} = \omega \wedge \star \omega = \left( -1 \right)^{|\omega| \left( m + n - |\omega| \right)} \star \omega \wedge \omega \ ,
\end{equation}
where $|\omega|$ indicates the parity of the form $\omega$. Notice that for supermanifolds that admit self-dual forms, we have that both the bosonic dimension and the fermionic one are even, thus it follows
\begin{equation}\label{ADHD}
	\mathpzc{B}er^{(n|m)} = \omega \wedge \star \omega = \left( -1 \right)^{|\omega|} \star \omega \wedge \omega \ .
\end{equation}
This leads to
\begin{equation}\label{ADHE}
	\star^2 \omega = (-1)^{|\omega|} \omega \ .
\end{equation}
In the usual bosonic setting, given a manifold  $\mathcal{M}^{(2n)}$, since in a given space $\Omega^{(p)} \left( \mathcal{M}^{(2n)} \right)$ all the forms have the same parity $|\omega | = p \mod 2 , \forall \omega \in \Omega^{(p)} \left( \mathcal{M}^{(2n)} \right)$, it follows from \eqref{ADHE} that $\star$ is an (anti-)involution of period 2, hence the operator splits the stable space $\displaystyle \Omega^{(n)} \left( \mathcal{M}^{(2n)} \right)$ in two parts, namely the self-dual part and the anti-self-dual part. In the super-setting this is not the case, since given a fixed space $\Omega^{(p|q)} \left( \mathcal{SM}^{(2m|2n)} \right)$ it contains both even forms and odd forms. If we impose the self-duality constraint on the stable space $\Omega^{(m|n)} \left( \mathcal{SM}^{(2m|2n)} \right)$ we have that the odd forms are identically set to 0:
\begin{equation}\label{ADHF}
	\star \omega = \omega \ \implies \ \star^2 \omega = \star \omega \ \implies \ \omega = 0 \ .
\end{equation}
Thus the constraint halves the dimension\footnote{This as to be interpreted carefully, recall that the spaces of pseudoforms are (countably) infinite-dimensional.} of the even forms and annihilates the odd forms. The same result hold for the anti-self-duality constraint. Thus, the net recipe for counting the degrees of freedom is: when considering a pseudoform $\omega \in \Omega^{(m|n)} \left( \mathcal{SM}^{(2m|2n)} \right)$, the (anti-)self duality constraint annihilates the odd forms and halves the number of even forms. This will be shown explicitly in the following.

Let us now move to the concrete analysis of self-dual pseudoforms on $\mathcal{SM}^{(2|2)}$. The only stable space (w.r.t. the action of the operator $\star$ defined in \eqref{ADHA}) is given by $\Omega^{(1|1)} \left( \mathcal{SM}^{(2|2)} \right)$. We can separate any pseudoform $Q^{(1|1)} \in \Omega^{(1|1)} \left( \mathcal{SM}^{(2|2)} \right)$ in a piece along $\delta \left( \psi \right)$ and a second one along $ \delta \left( \bar{\psi} \right)$, thus we write
\begin{eqnarray}\label{SDFPB}
	Q^{(1|1)} = Q^{(1|1)}_+ + Q^{(1|1)}_- \ ,
\end{eqnarray}
where in particular (we will sometimes use the shorter notations $\delta \equiv \delta \left( \psi \right)$, $\bar{ \delta} = \delta \left( \bar{\psi} \right)$, $\delta^{(n)} \equiv \iota^n \delta \left( \psi \right)$ and $\bar{\delta}^{(n)} \equiv \bar{\iota}^n \delta \left( \bar{\psi} \right)$)
\begin{eqnarray}
	\nonumber Q^{(1|1)}_+ &=& \sum_{n=0}^\infty \left[ Q_{+,0,n} \bar{\psi}^{n+1} \delta^{(n)} + Q_{+,V,n} V \bar{\psi}^{n} \delta^{(n)} + Q_{+,\bar{V},n} \bar{V} \bar{\psi}^{n} \delta^{(n)} + Q_{+,V \bar{V},n} V \bar{V} \bar{\psi}^{n} \delta^{(n+1)} \right] = \\
	\label{SDFPBA} && = \sum_{n=0}^\infty \bar{\psi}^n \iota^n \left[ Q_{+,0,n} \bar{\psi} + Q_{+,V,n} V + Q_{+,\bar{V},n} \bar{V} + Q_{+,V \bar{V},n} V \bar{V} \iota \right] \delta \left( \psi \right) \ \ ; \\
	\nonumber Q^{(1|1)}_- &=& \sum_{n=0}^\infty \left[ Q_{-,0,n}\psi^{n+1} \bar{\delta}^{(n)} + Q_{-,V,n} V \psi^{n} \bar{\delta}^{(n)} + Q_{-,\bar{V},n} \bar{V} \psi^{n} \bar{\delta}^{(n)} + Q_{-,V \bar{V},n} V \bar{V} \psi^{n} \bar{\delta}^{(n+1)} \right] = \\
	\label{SDFPBB} && = \sum_{n=0}^\infty \psi^n \bar{\iota}^n \left[ Q_{-,0,n} \psi + Q_{-,V,n} V + Q_{-,\bar{V},n} \bar{V} + Q_{-,V \bar{V},n} V \bar{V} \bar{\iota} \right] \delta \left( \bar{\psi} \right) \ \ .
\end{eqnarray}
Actually, we want now to analyse the self-duality prescription in order to construct an integral form $Q^{(1|2)}$ (as the one appearing in \eqref{streAA}) which is given by
\begin{equation}\label{SDFPA}
	Q^{(1|2)} = Q^{(1|1)} \wedge \left( \mathbb{Y}^{(0|1)} + \bar{\mathbb{Y}}^{(0|1)} \right) \ ,
\end{equation}
where $\mathbb{Y}^{(0|1)} = V \iota \delta \left( \psi \right)$ and $\bar{\mathbb{Y}}^{(0|1)} = \bar{V} \bar{\iota} \delta \left( \bar{\psi} \right)$, while $Q^{(1|1)}$ is a self-dual pseudoform. Notice that, a priori, one could consider any linear combination of the two PCO's; here we will consider only the case shown in \eqref{SDFPA} corresponding to a real PCO. For the sake of clarity, we are using the supersymmetric PCO's, while a priori one could think to use different PCO's, e.g. the component ones. In the next section we will give a brief mention to the choice of components operators, showing that actually one comes to analogous results. This is again far from trivial, since when introducing terms à la SFT one looses control with the notions of \virgolette closure" of the Lagrangian, since in this case we will have the rheonomic Lagrangian represented by a superform and the SFT term represented by a pseudoform. \\

It is easy to verify that all the terms containing $\psi^2, \bar{\psi}^2, \iota^2$ or $\bar{\iota}^2$ vanish identically when inserted in \eqref{SDFPA}, thus we do not lose in generality if we consider a pseudoform built as follows (recall $\delta' \equiv \iota \delta$, the same for $\bar{\delta}'$):
\begin{eqnarray}
	\label{SDFPC} Q^{(1|1)}_- &=& Q_1 \psi \bar{\delta} + Q_2 V \bar{\delta} + Q_3 \bar{V} \bar{\delta} + Q_4 V \psi \bar{\delta}' + Q_5 \bar{V} \psi \bar{\delta}' + Q_6 V \bar{V} \bar{\delta}' \ , \\
	\nonumber Q^{(1|1)}_+ &=& Q_7 \bar{\psi} \delta + Q_8 V \delta + Q_9 \bar{V} \delta + Q_{10} V \bar{\psi} \delta' + Q_{11} \bar{V} \bar{\psi} \delta' + Q_{12} V \bar{V} \delta' \ .
\end{eqnarray}

Before giving the final expression of the self-dual pseudoform built from \eqref{SDFPC}, and its corresponding integral form given by \eqref{SDFPA}, let us list some useful expressions derived from \eqref{ADHA}:
\begin{align}
	\nonumber & \star \left[ \psi \delta \left( \bar{\psi} \right) \right] = V \bar{V} \iota \delta \left( \psi \right) \ \ , \ \ \star \left[ \bar{\psi} \delta \left( \psi \right) \right] = - V \bar{V} \bar{\iota} \delta \left( \bar{\psi} \right) \ \ , \ \ \star \left[ V \delta \left( \psi \right) \right] = - \bar{V} \delta \left( \bar{\psi} \right) \ \ , \\
	\nonumber & \star \left[ V \delta \left( \bar{\psi} \right) \right] = \bar{V} \delta \left( \psi \right) \ \ , \ \ \star \left[ \bar{V} \delta \left( \psi \right) \right] = V \delta \left( \bar{\psi} \right) \ \ , \ \ \star \left[ \bar{V} \delta \left( \bar{\psi} \right) \right] = - V \delta \left( \psi \right) \ \ , \\
	\nonumber & \star \left[ V \psi \bar{\iota} \delta \left( \bar{\psi} \right) \right] = \bar{V} \bar{\psi} \iota \delta \left( \psi \right) \ \ , \ \ \star \left[ \bar{V} \psi \bar{\iota} \delta \left( \bar{\psi} \right) \right] = - V \bar{\psi} \iota \delta \left( \psi \right) \ \ , \ \ \star \left[ V \bar{\psi} \iota \delta \left( \psi \right) \right] = - \bar{V} \psi \bar{\iota} \delta \left( \bar{\psi} \right) \ \ , \\
	\label{ADHB} & \star \left[ \bar{V} \bar{\psi} \iota \delta \left( \psi \right) \right] = V \psi \bar{\iota} \delta \left( \bar{\psi} \right) \ \ , \ \ \star \left[ V \bar{V} \bar{\iota} \delta \left( \bar{\psi} \right) \right] = \bar{\psi} \delta \left( \psi \right) \ \ , \ \ \star \left[ V \bar{V} \iota \delta \left( \psi \right) \right] = - \psi \delta \left( \bar{\psi} \right) \ \ .
\end{align}
These expressions allow us to implement the self-duality prescription that hence leads to
\begin{eqnarray}\label{SDFPE}
	Q^{(1|1)} = Q_2 \left( V \bar{\delta} + \bar{V} \delta \right) + Q_3 \left( \bar{V} \bar{\delta} - V \delta \right) + Q_4 \left( V \psi \bar{\delta}' + \bar{V} \bar{\psi} \delta' \right) + Q_5 \left( \bar{V} \psi \bar{\delta}' - V \bar{\psi} \delta' \right) \ . 
\end{eqnarray}
Notice in particular that, as we discussed previously, the terms corresponding to odd forms have been eliminated, while the degrees of freedom of the terms corresponding to bosonic forms have been halved. From this self-dual pseudoform we can generate the corresponding integral form as in \eqref{SDFPA}:
\begin{equation}\label{SDFPF}
	Q^{(1|2)} = Q^{(1|1)} \left( \mathbb{Y}^{(0|1)} + \bar{\mathbb{Y}}^{(0|1)} \right) = \left( Q_3 + Q_5 \right) \left( V \bar{V} \delta \bar{\delta}' - V \bar{V} \delta' \bar{\delta} \right) \ .
\end{equation}
The pieces containing $Q_2$ or $Q_4$ are annihilated both by $\mathbb{Y}^{(0|1)}$ and $\bar{\mathbb{Y}}^{(0|1)}$.

It is worthy to consider the inverse problem, i.e., whether given an integral form $Q^{(1|2)}$, one can reconstruct the corresponding self-dual pseudoform. In other words, we are asking if there is a way to give a consistent \virgolette self-duality prescription" on an integral form.\footnote{Clearly, the resulting integral form will not be self-dual in the usual sense, we are looking for an integral form originated by a self-dual pseudoform.} Let us then consider a generic $(1|2)$ integral form
\begin{equation}\label{SDFPG}
	\tilde{Q}^{(1|2)} = A V \bar{V} \delta' \bar{\delta} + B V \bar{V} \delta \bar{\delta}' + C V \delta \bar{\delta} + D \bar{V} \delta \bar{\delta} \ .
\end{equation}
It is possible to obtain a self-dual pseudoform that generates $\tilde{Q}^{(1|2)}$ by using the new PCO's
\begin{equation}\label{SDFPH}
	\mathbb{Y}^{(0|-1)\dagger} = \star \mathbb{Y}^{(0|1)} \star \equiv \mathbb{Y}^\dagger \ \ , \ \ \bar{\mathbb{Y}}^{(0|-1)\dagger} = \star \bar{\mathbb{Y}}^{(0|-1)} \star \equiv \bar{\mathbb{Y}}^\dagger \ .
\end{equation}
Really, we can apply the PCO $\tilde{\mathbb{Y}}^{\dagger} = \alpha \mathbb{Y}^{\dagger} + \beta \bar{\mathbb{Y}}^{\dagger}$, with $\alpha , \beta$ generic coefficients:
\begin{eqnarray}
	\nonumber && \star \left( A V \bar{V} \delta' \bar{\delta} + B V \bar{V} \delta \bar{\delta}' + C V \delta \bar{\delta} + D \bar{V} \delta \bar{\delta} \right) = - A \psi - B \bar{\psi} + C \bar{V} - D V \ \ ; \\
	\nonumber && \left( - A \psi - B \bar{\psi} + C \bar{V} - D V \right) \wedge \left( \alpha \mathbb{Y}^{(0|1)} + \beta \bar{\mathbb{Y}}^{(0|1)} \right) = \\
	\nonumber && = \alpha A V \delta - \beta A \bar{V} \psi \bar{\delta}' - \alpha B V \bar{\psi} \delta' + \beta B \bar{V} \delta - C V \bar{V} \iota \delta - D V \bar{V} \bar{\iota} \bar{\delta} \ \ ; \\
	\label{SDFPI} && \star \left( \alpha \mathbb{Y}^{(0|1)} + \beta \bar{\mathbb{Y}}^{(0|1)} \right) \star Q^{(1|2)} = - \alpha A \bar{V} \bar{\delta} + \beta A V \bar{\psi} \delta' + \alpha B \bar{V} \psi \bar{\delta}' - \beta B V \delta + C \psi \bar{\delta} - D \bar{\psi} \delta \ \ .
\end{eqnarray}
By confronting \eqref{SDFPI} with \eqref{SDFPE} we obtain
\begin{equation}\label{SDFPJ}
	Q_3 = - \alpha A \ \ , \ \ Q_5 = \beta B \ \ , \ \ Q_3 = \alpha B \ \ , \ \ Q_5 = - \beta A \ \ , \ \ C = 0 \ \ , \ \ D = 0 \ \  \implies $$ $$ \implies \ \  A = - \frac{\beta}{\alpha} B \ \ , \ \ A = - \frac{\alpha}{\beta} B \ \ \implies \ \ \alpha^2 = \beta^2 \ .
\end{equation}
Notice that, as expected, we could have removed any piece which is either in the kernel of $\mathbb{Y}^{(0|1)}$ or in that of $\bar{\mathbb{Y}}^{(0|1)}$, i.e., the terms containing either $C$ or $D$. For the sake of clarity we can fix $\alpha = \beta = 1$ (corresponding again to the real PCO), so that it is possible to find the \virgolette originating" pseudoform if $A = - B$.

Hence, the self-duality prescription on integral forms reads
\begin{equation}\label{SDFPL}
	\star \tilde{\mathbb{Y}}^{\dagger} Q^{(1|2)} = \tilde{\mathbb{Y}}^{\dagger} Q^{(1|2)} \ ,
\end{equation}
leading to
\begin{equation}\label{SDFPP}
	Q^{(1|2)} = B \left( V \bar{V} \delta \bar{\delta}' - V \bar{V} \delta' \bar{\delta} \right) \ ,
\end{equation}
which is exactly the same form as \eqref{SDFPF}. This shows that both if we start from a self-dual pseudoform and lift it to an integral form and if we start from an integral form and apply to it a self-duality prescription we come to the same result.

Let us now discuss the analogous result carried on with the component PCO. In particular, let us consider \eqref{SDFPA} with the PCO's $\mathbb{Y} = \theta \delta$, $\bar{\mathbb{Y}} = \bar{\theta} \bar{\delta}$:
\begin{equation}\label{SDFPQ}
	Q^{(1|2)} = Q^{(1|1)} \wedge \left( \theta \delta + \bar{ \theta} \bar{\delta} \right) \ \ , \ \ Q^{(1|1)} = \star Q^{(1|1)} \ \ .
\end{equation}
In this case, we see from \eqref{SDFPBA} and \eqref{SDFPBB} that only the following terms should be considered (the others are trivially 0 after the projection with the component PCO's):
\begin{equation}\label{SDFPQA}
	Q^{(1|1)} = Q_1 V \delta + Q_2 \bar{V} \delta + Q_3 V \bar{V} \iota \delta + Q_4 V \bar{\delta} + Q_5 \bar{V} \bar{\delta} + Q_6 V \bar{V} \bar{\iota} \bar{\delta} \ .
\end{equation}
By imposing the self-duality constraint we obtain the most general self-dual pseudoform that fits into \eqref{SDFPQ}:
\begin{equation}\label{SDFPR}
	Q^{(1|1)} = Q_1 \left( V \bar{\delta} + \bar{V} \delta \right) + Q_2 \left( V \delta - \bar{V} \bar{\delta} \right) \ \ .
\end{equation}
By inserting \eqref{SDFPR} in \eqref{SDFPQ} we obtain
\begin{equation}\label{SDFPS}
	Q^{(1|2)} = \left( - Q_1 \theta - Q_2 \bar{\theta} \right) V \delta \bar{\delta} + \left( Q_1 \bar{\theta} - Q_2 \theta \right) \bar{V} \delta \bar{\delta} \ \ .
\end{equation}
A priori it may seem that this result is not compatible with the one we found in \eqref{SDFPF}, but indeed this is not true. In order to see this explicitly, let us count the number of degrees of freedom in the two cases: let us suppose the following decomposition of a superfield $Q_i$:
\begin{equation}\label{SDFPSA}
	Q_i = q_i + \theta q_{i \theta} + \bar{\theta} q_{i \theta} + \theta \bar{\theta} q_{i \theta \bar{\theta}} \ .
\end{equation}
In \eqref{SDFPF} we have four degrees of freedom, since the two superfields appear only in the combination $Q_3 + Q_5$; in \eqref{SDFPS} we have that
\begin{eqnarray}
	\label{SDFPSB} && - Q_1 \theta - Q_2 \bar{\theta} = - q_1 \theta - q_2 \bar{\theta} + \left[ \left( -1 \right)^{|q_{2\theta}|} q_{2\theta} - \left( -1 \right)^{|q_{1 \bar{\theta}}|} q_{1 \bar{\theta}} \right] \theta \bar{\theta} \ , \\
	\label{SDFPSC} && Q_1 \bar{\theta} - Q_2 \theta = q_1 \bar{\theta} - q_2 \theta + \left[ \left( -1 \right)^{|q_{1 \theta}|} q_{1 \theta} + \left( -1 \right)^{|q_{2 \bar{\theta}}|} q_{2 \bar{\theta}} \right] \theta \bar{\theta} \ .
\end{eqnarray}
From \eqref{SDFPSB} and \eqref{SDFPSC} we count again four degrees of freedom, two bosonic and two fermionic as in the previous case.

After this preamble on self-dual forms on $\mathcal{SM}^{(2|2)}$ we are now ready to discuss the action for the chiral boson.

\section{The Action: Coupling with Self-Dual Pseudoform}

In the previous section we have given a consistent prescription for the self-duality condition on the supermanifold $\mathcal{SM}^{(2|2)}$. In this section we explicitly calculate the action defined in \eqref{streAA} (we also add the coupling with external fields) for the supersymmetric and the component PCO's, in order to show which degrees of freedom decouple from the theory and which ones propagate.

\subsection{Supersymmetric PCO}

By following the prescriptions given in the introduction, the action reads
\begin{eqnarray}
	\label{TAWP0} S = \int_{\mathcal{SM}^{(2|2)}} \Big[ && \mathcal{L}^{(2|0)}_c \wedge \mathbb{Y}^{(0|2)} + d \Phi^{(0|0)} \wedge Q^{(1|2)} + f^{(2|2)} \left( M,Q \right) \Big] \ ,
\end{eqnarray}
where the last term represents the coupling of the form $Q^{(1|2)}$ with the external fields generically denoted $M$. The first term contains the PCO $\mathbb{Y}^{(0|2)}$ which we now fix to $\mathbb{Y}^{(0|2)} = V \iota \delta \left( \psi \right) \bar{V} \bar{\iota} \delta \left( \bar{\psi} \right)$. In this case, the first term of \eqref{TAWP0} involves the superfields $\Phi$ and $W$ only:
\begin{equation}\label{TAWP0A}
	\mathcal{L}^{(2|0)}_c \wedge \mathbb{Y}^{(0|2)} = \bar{D} \Phi W V \bar{V} \delta \left( \psi \right) \delta \left( \bar{\psi} \right) \ .
\end{equation}

Let us start by calculating the term $\displaystyle d \Phi^{(0|0)} \wedge Q^{(1|2)} $, with $Q^{(1|2)}$ given as as in \eqref{SDFPF} (we define $Q_3 + Q_5 = \Lambda$):
\begin{align}\label{TAWPA}
	d \Phi^{(0|0)} \wedge Q^{(1|2)} = \left[ - \bar{D} \Phi + D \Phi \right] \Lambda V \bar{V} \delta \left( \psi \right) \delta \left( \bar{\psi} \right) \ .
\end{align}
Notice that, in order to respect the parity of the action, $\Lambda$ is an odd superfield. By inserting \eqref{TAWPA} in \eqref{TAWP0}, we obtain
\begin{eqnarray}\label{TAWPC}
	\nonumber S = \int_{\mathcal{SM}^{(2|2)}} \left[ \left( \bar{D} \Phi W - \bar{D} \Phi \Lambda + D \Phi \Lambda \right) V \bar{V} \delta \left( \psi \right) \delta \left( \bar{\psi} \right) + f^{(2|2)} (M,\mathcal{Q}) \right] \ .
\end{eqnarray}
The term containing the coupling with external fields $M$ is a $(2|2)$-integral form $f^{(2|2)}(M,Q)$; since $Q^{(1|2)}$ is a picture-2 form, we have that $f^{(2|2)}(M,Q)$ is at most linear in $Q^{(1|2)}$, hence it generically reads
\begin{equation}\label{TAWPD}
	f^{(2|2)}(M,Q) = R^{(1|0)}(M) \wedge Q^{(1|2)} + f^{(2|2)} (M) = \left[ \left( R_\psi (M) - R_{\bar{\psi}} (M) \right) \Lambda + f(M) \right] V \bar{V} \delta \bar{\delta} \ .
\end{equation}
Then the full action reads
\begin{equation}\label{TAWPE}
	S = \int_{z , \bar{z} , \theta , \bar{\theta}} \hspace{-.7cm} \left( \bar{D} \Phi W - \bar{D} \Phi \Lambda + D \Phi \Lambda + \left( R_\psi - R_{\bar{\psi}} \right) \Lambda + f (M) \right) \ .
\end{equation}
The equations of motion are
\begin{eqnarray}
	\label{TAWPF} \bar{D} \Phi = 0 \ \ , \ \ - \bar{D} W + \bar{D} \Lambda - D \Lambda = 0 \ \ , \\
	\label{TAWPG} \bar{D} \Phi - D \Phi - R_\psi + R_{\bar{\psi}} = 0 \ \ , \\
	\label{TAWPH} \frac{\delta S}{\delta M} = 0 \ \ .
\end{eqnarray}
Notice that, as expected, the equations \eqref{TAWPF} imply that only the $\bar{D}$ part of the superfield $W$ couples to the superfield $\Lambda$. We can act on \eqref{TAWPG} with the superderivative $\bar{D}$ and then substitute the expression for $\bar{D} \Phi$ from \eqref{TAWPF} in order to obtain
\begin{eqnarray}
	- \bar{D} R_\psi + \bar{D} R_{\bar{\psi}} = 0 \ .
\end{eqnarray}
This equation together with \eqref{TAWPH} provide an independent set of equations for $\Lambda$ and $M$. Given one of their solutions we can compute $W$ and $\Phi$ from \eqref{TAWPG} and \eqref{TAWPF}. The key point is that different solutions to the previous equations differ from each other by free field equations of motion
\begin{eqnarray}\label{TAWPJ}
	\bar{D} \Delta \Phi = 0 \ \ , \ \ D \Delta \Phi = 0 \ \ , \ \ \bar{D} \Delta W = 0 \ \ .
\end{eqnarray}
The first two equations imply constant fluctuations $\Delta \Phi$ of $\Phi$, which can be discarded (up to zero-mode counting), the last equation of \eqref{TAWPJ} yields holomorphic fluctuations $\Delta W$ of $W$, whose component content amounts to a chiral boson and a chiral fermion.

This argument shows that one needs to use the full power of supergeometry (in particular the full complex of differential forms) in order to keep the geometrical point of view of a field theory under control. In this case we have shown that pseudoforms are essential objects in order to overcome some limitations that rheonomy has built-in.

\subsection{Component PCO}

In this subsection we want to show that we can obtain the same results of the previous one by considering the \virgolette self-dual" form $Q^{(1|2)}$ as in \eqref{SDFPS}. The coupling term $d \Phi^{(0|0)} \wedge Q^{(1|2)}$ reads
\begin{equation}\label{TAWPQ}
	d \Phi^{(0|0)} \wedge Q^{(1|2)} = \left[ \left( - Q_1 \theta - Q_2 \bar{\theta} \right) \bar{\partial} \Phi + \left( -Q_1 \bar{\theta} + Q_2 \theta \right) \partial \Phi \right] V \bar{V} \delta \bar{\delta} \ \ .
\end{equation}
Analogously, the term containing the other fields $R^{(1|0)} \wedge Q^{(1|2)}$ reads
\begin{equation}\label{TAWPR}
	R^{(1|0)} \wedge Q^{(1|2)} = \left[ \left( - Q_1 \theta - Q_2 \bar{\theta} \right) R_{\bar{V}} + \left( - Q_1 \bar{\theta} + Q_2 \theta \right) R_V \right] V \bar{V} \delta \bar{\delta} \ \ .
\end{equation}
The action, already expanded in components, reads
\begin{align}
	\nonumber S = \int_{_{z \bar{z}}} &\Big[ \phi_{\bar{\theta}} w_{\theta \bar{\theta}} - \phi_{\theta \bar{\theta}} w_{\bar{\theta}} + \bar{\partial} \phi w_{\theta} + \bar{\partial} \phi_{\theta} w - q_{1} \bar{\partial} \phi_{\bar{\theta}} + \left( q_{2,\theta} - q_{1, \bar{\theta}} \right) \bar{\partial} \phi + q_{2} \bar{\partial} \phi_\theta + q_{1} \partial \phi_\theta + \\
	\nonumber &- \left( q_{1,\theta} + q_{2 \bar{\theta}} \right) \partial \phi + q_{2} \partial \phi_{\bar{\theta}} - q_{1} R_{\bar{V},\bar{\theta}} + \left( q_{2,\theta} - q_{1, \bar{\theta}} \right) R_{\bar{V}, 0} + q_{2} R_{\bar{V}, \theta} + q_{1} R_{V,\theta} + \\
	\label{TAWPS} &- \left( q_{1,\theta} + q_{2,\bar{\theta}} \right) R_{V,0} + q_{2} R_{V, \bar{\theta}} + f \Big] \ \ .
\end{align}
We can rearrange the six fields emerging from $Q_{1/2}$ in the four independent combinations as follows:
\begin{equation}\label{TAWPT}
	-q_{1} = \eta_1 \ , \ q_{2} = \eta_2 \ , \ q_{2,\theta} - q_{1, \bar{\theta}} = p_1 \ , \ -q_{1,\theta} - q_{2, \bar{\theta}} = p_2 \ \ .
\end{equation}
Then the action reads
\begin{align}
	\nonumber S = \int_{_{z \bar{z}}} &\Big[ \phi_{\bar{\theta}} w_{\theta \bar{\theta}} - \phi_{\theta \bar{\theta}} w_{\bar{\theta}} + \bar{\partial} \phi w_{\theta} + \bar{\partial} \phi_{\theta} w + \eta_1 \left( \bar{\partial} \phi_{\bar{\theta}} - \partial \phi_\theta + R_{\bar{V}, \bar{\theta}} - R_{V, \theta} \right) + p_1 \left( \bar{\partial} \phi + R_{\bar{V},0} \right) + \\
	\label{TAWPU} &+ \eta_2 \left( \bar{\partial} \phi_\theta + \partial \phi_{\bar{\theta}} + R_{\bar{V},\theta} + R_{V, \bar{\theta}} \right) + p_2 \left( \partial \phi + R_{V,0} \right) + f \Big] \ \ .
\end{align}
The equations of motion can be easily obtained:
\begin{eqnarray}
	\label{TAWPV} \frac{\delta S}{\delta W} = 0 \ &\implies& \ \bar{\partial} \phi_{\theta} = 0 \ , \ \bar{\partial} \phi = 0 \ , \ \phi_{\theta \bar{\theta}} = 0 \ , \ \phi_{\bar{\theta}} = 0 \ \ , \\
	\nonumber \frac{\delta S}{\delta \Phi} = 0 \ &\implies& \ - \bar{\partial} w_\theta - \bar{\partial} p_1 - \partial p_2 = 0 \ , \ - \bar{\partial} w - \partial \eta_1 + \bar{\partial} \eta_2 = 0 \ \ , \\
	\label{TAWPW} && w_{\theta \bar{\theta}} + \bar{\partial} \eta_1 + \partial \eta_2 = 0 \ , \ w_{\bar{\theta}} = 0 \ \ , \\
	\label{TAWPX} \frac{\delta S}{\delta \eta_{_{1/2}}} = 0 \ &\implies& \ \bar{\partial} \phi_{\bar{\theta}} - \partial \phi_\theta + R_{\bar{V}, \bar{\theta}} - R_{V, \theta} = 0 \ , \ \bar{\partial} \phi_\theta + \partial \phi_{\bar{\theta}} + R_{\bar{V},\theta} + R_{V, \bar{\theta}} = 0 \ \ , \\
	\label{TAWPY} \frac{\delta S}{\delta p_{_{1/2}}} = 0 \ &\implies& \ \bar{\partial} \phi + R_{\bar{V},0} = 0 \ , \ \partial \phi + R_{V,0} = 0 \ \ , \\
	\label{TAWPZ} \frac{\delta S}{\delta M} = 0 \ \ .
\end{eqnarray}
We can use \eqref{TAWPV} in \eqref{TAWPX} and \eqref{TAWPY} (and in case apply $\bar{\partial}$) in order to obtain equations for $R$ (namely, for $M$):
\begin{equation}\label{TAWPZA}
	\bar{\partial} R_{\bar{V}, \bar{\theta}} - \bar{\partial} R_{V, \theta} = 0 \ , \ R_{\bar{V},\theta} + R_{V, \bar{\theta}} = 0 \ , \ R_{\bar{V},0} = 0 \ , \ \bar{\partial} R_{V,0} = 0 \ \ .
\end{equation}
We then see that in this case $\phi$ and $\phi_\theta$ are the decoupled chiral boson and chiral fermions respectively, while from \eqref{TAWPW} we easily see that only the chiral part of $w$ (fermion) and $w_\theta$ (boson) couple to the rest of the fields, in exact analogy with the previous case. This confirms once again that the results obtained with the supersymmetric and the component PCO's are consistent. 

\subsection{Non-Factorised Action: an Alternative Approach?}

In the previous calculations, there might be a caveat: the rheonomic Lagrangian built in this way is still not closed. We have shown that we get the correct counting of degrees of freedom by using the supersymmetric PCO. This fact suggests that it may exist a more general formulation. Once again, SFT suggests a way: is it possible to lift the rheonomic Lagrangian to a non-factorised one? One should then lift the fields in the rheonomic Lagrangian to genuine pseudoforms, so that the picture number needs not to be saturated via PCO's. This approach has been investigated in \cite{Cremonini:2019aao} in the context of super Chern-Simons theory: instead of using factorised fields $A^{(1|0)} \wedge \mathbb{Y}^{(0|1)}$, the authors used pseudoforms $A^{(1|1)}$. This lead to the emerging of non-trivial algebraic structures ($A_\infty$ and $L_\infty$ algebras) which mimic the same structures of open Superstring Field Theory (see, e.g., the recent \cite{Erler:2013xta}).

\noindent We see that if we consider the non-factorised action, the SFT-inspired term is already built-in. Indeed, we show that a simple field redefinition produces the term $d \Phi \wedge Q$.

Let us \virgolette lift" the fields in \eqref{oneQ} to picture 1 fields, so that the resulting action reads
\begin{equation}
	S = \int_{\mathcal{SM}^{(2|2)}} \Big[(\xi^{(0|1)} V + \bar\xi^{(0|1)} \bar V) \wedge (d\Phi^{(0|1)} - \psi W^{(0|1)}) + \xi^{(0|1)} \bar \xi^{(0|1)} V\wedge \bar V + $$ $$ + W^{(0|1)} d W^{(0|1)} \wedge V -  d\Phi^{(0|1)} \wedge \psi W^{(0|1)} \Big] \ .
\end{equation}
We wonder if there exist a field redefinition that generates Sen's term. In order to verify this, we can rewrite the action as
\begin{equation}
	S = \int_{\mathcal{SM}^{(2|2)}} \Big[ (\xi^{(0|1)} V + \bar\xi^{(0|1)} \bar V + \psi W^{(0|1)}) d\Phi^{(0|1)} - \psi W^{(0|1)} \left( \xi^{(0|1)} V + \bar\xi^{(0|1)} \bar V \right) + $$ $$ + \xi^{(0|1)} \bar \xi^{(0|1)} V\wedge \bar V + 
W^{(0|1)} d W^{(0|1)} \wedge V \Big] \ .
\end{equation}
Let us consider the first term of the action and denote it as
\begin{equation}
	A^{(1|1)} \wedge d\Phi^{(0|1)} \ , \ A^{(1|1)} = (\xi^{(0|1)} V + \bar\xi^{(0|1)} \bar V + \psi W^{(0|1)}) \ .
\end{equation}
We can manipulate this expression as
\begin{equation}
	A^{(1|1)} = A^{(1|1)} + \star A^{(1|1)} - \star A^{(1|1)} = Q^{(1|1)} - \star A^{(1|1)} \ ,
\end{equation}
where $Q$ is self-dual. Actually one should pay attention at this point: as we noticed in the previous sections, $\star^2$ produces a minus sign if the form on which it acts is odd. We assume that one correctly groups together the even terms from $\star A$ and the odd terms from $-\star A$ and eventually field-redefines the remaining fields to correct signs. Here we are are skipping these details, we just want to give an intuition of the emergence of the $Q$ term. We have then obtained\vspace{-0.3cm}
\begin{equation}
	A^{(1|1)} \wedge d\Phi^{(0|1)} = Q^{(1|1)} \wedge d\Phi^{(0|1)} - \star A^{(1|1)} \wedge d\Phi^{(0|1)} = Q^{(1|1)} \wedge d\Phi^{(0|1)} - A^{(1|1)} \wedge \star d\Phi^{(0|1)} \ ,
\end{equation}
where we have moved the Hodge operator by implicitly assuming the integration over $\mathcal{SM}^{(2|2)}$. We can now re-define $\displaystyle - \star d\Phi^{(0|1)} = \left( d\Phi^{(0|1)} \right)' $ (notice that $d \Phi$ is a $(1|1)$-form) and the term with $Q$ reads
\begin{equation}
	Q^{(1|1)} \wedge d\Phi^{(0|1)} = \star Q^{(1|1)} \wedge d\Phi^{(0|1)} = Q^{(1|1)} \wedge \star d\Phi^{(0|1)} = - Q^{(1|1)} \wedge \left( d\Phi^{(0|1)} \right)' = \left( d\Phi^{(0|1)} \right)' \wedge Q^{(1|1)} \ .
\end{equation}
The field re-defined action now reads
\begin{equation}
	S = \int_{\mathcal{SM}^{(2|2)}} \Big[ (\xi^{(0|1)} V + \bar\xi^{(0|1)} \bar V + \psi W^{(0|1)}) \left( d\Phi^{(0|1)} \right)' - \psi W^{(0|1)} \left( \xi^{(0|1)} V + \bar\xi^{(0|1)} \bar V \right) + $$ $$ + \xi^{(0|1)} \bar \xi^{(0|1)} V\wedge \bar V + 
W^{(0|1)} d W^{(0|1)} \wedge V + \left( d\Phi^{(0|1)} \right)' \wedge Q^{(1|1)} \Big] \ ,
\end{equation}
that is, the action with Sen's term inserted. This argument shows (or less presumptuously seems to show) that, when working with pseudoforms, Sen's term is already included. This resembles what we have found in \cite{Cremonini:2019aao,Cremonini:2019xco} for the supersymmetric term of super Chern-Simons theory: when working with the non-factorised Lagrangian, it is not necessary to add other terms in order to obtain a supersymmetric action, since it is supersymmetric by construction.

As we said, this method has not been investigated yet and it may carry other challenges; for example,  pseudoforms have an infinite number of components (for super Chern-Simons \cite{Cremonini:2019aao,Cremonini:2019xco} we have shown how to deal with this problem). Are pseudoforms a new way to deal with auxiliary fields?

\section*{Conclusions}\label{sec:conclusions}

\addcontentsline{toc}{section}{\nameref{sec:conclusions}}

\noindent The present work opens new directions in the study of supersymmetric models, supergravity and 
string theory, by using the powerful techniques of supergeometry. It gives a path from string field theory to quantum field 
theories, justifying the cited results and leading to several generalizations such as non-flat backgrounds and higher dimensional models 
(for example, $M5$-branes \cite{Lambert:2019khh,Andriolo:2020ykk}, supergravity couplings, string theory on supermanifolds). 
The old problem of self-dual field strengths and the absence of 
auxiliary fields might be faces of the same coin and we believe that they both can be solved by introducing new non-factorized couplings
of the Sen's form. The technique introduced here might have a relevant impact in future analysis.  The underlying string field 
theory and the powerfulness of supergeometric techniques provide a universal constructive method. We plan to develop this program for the $M5$-branes studied from a supergeometrical point of view in a forthcoming publication \cite{prep}.

\section*{Acknowledgements}
\noindent This work has been partially supported by Universit\`a del Piemonte Orientale research funds. We thank L. Castellani, R. Catenacci and S. Noja for many useful discussions. We also thank A. Sen for his comments on the draft.

\appendix

\section{A Brief Review on Integral Forms}

In this appendix we want to recall the main definitions and computation techniques for integral forms. For more exhaustive or more rigorous approaches to integral forms the reader to, e.g., \cite{Belo,Witten:2012bg,Castellani:2015paa,Catenacci:2018xsv,Cremonini:2019aao}.

We consider a supermanifold ${\cal SM}^{(n|m)}$ with $n$ bosonic and $m$ fermionic dimensions. We denote the local coordinates in an open set as $(x^a, \theta^\alpha), a=1,\ldots,n , \alpha=1,\ldots,m$. A generic $(p|0)$-form, i.e., a \emph{superform}, has the following local expression
\begin{equation}\label{ABRIFA}
	\omega^{(p|0)} = \omega_{[i_1 \ldots i_r](\alpha_1 \ldots \alpha_s)} \left( x , \theta \right) dx^{i_1} \wedge \ldots \wedge dx^{i_r} \wedge d \theta^{\alpha_1} \wedge \ldots \wedge d \theta^{\alpha_s} \ , \ p=r+s \ .
\end{equation}
The coefficients $\omega_{[i_1 \ldots i_r](\alpha_1 \ldots \alpha_s)}(x,\theta)$ are a set of superfields and the indices $a_1 \dots a_r$, $\alpha_1 \dots \alpha_s$ are anti-symmetrized and symmetrised, respectively, because of the rules (we omit the \virgolette $\wedge$" symbol)
\begin{equation}\label{ABRIFB}
	dx^i dx^j = - dx^j dx^i \ , \ d \theta^\alpha d \theta^\beta = d \theta^\beta d \theta^\alpha \ , \ dx^i d \theta^\alpha = d \theta^\alpha d x^i \ .
\end{equation}
Namely, we assign \emph{parity} $1 \mod 2$ to odd forms and $0 \mod 2$ to even forms:
\begin{equation}\label{ABRIFC}
	\left| dx \right| = 1 \mod 2 \ , \ \left| d \theta \right| = 0 \mod 2 \ .
\end{equation}
Since superforms are generated also by commuting generators, we immediately see that there is not notion of top form among superforms. In other words, if one looks for the analogous of the Determinant bundle on a supermanifold, one has to consider a different class of forms: \emph{integral forms}. A generic integral form locally reads (see the list of references given at the beginning of the appendix for a mathematical introduction to integral forms and some physical applications)
\begin{equation}\label{ABRIFD}
	\omega^{(p|m)} = \omega_{[i_1 \ldots i_r]}^{(\alpha_1 \ldots \alpha_s)} \left( x , \theta \right) dx^{i_1} \wedge \ldots \wedge dx^{i_r} \wedge \iota_{\alpha_1} \ldots \iota_{\alpha_s} \delta \left( d \theta^1 \right) \wedge \ldots \wedge \delta \left( d \theta^m \right) \ ,
\end{equation}
where $\delta \left( d \theta \right)$ is a (formal) Dirac distribution, i.e., an object used to introduce an algebraic integration along even forms as we are going to see in a while, and $\iota_\alpha$ is the contraction operator along the (odd) vector field $\partial_\alpha$. These objects satisfy the following properties:
\begin{equation}\label{ABRIFE}
	\int_{d \theta} \delta \left( d \theta \right) = 1 \ \ , \ \ d \theta \delta \left( d \theta \right) = 0 \ \ , \ \ \delta \Big( d \theta^\alpha \Big) \wedge \delta \left( d \theta^\beta \right) = - \delta \left( d \theta^\beta \right) \wedge \delta \Big( d \theta^\alpha \Big) \ \ , \ \ dx \wedge \delta \left( d \theta \right) = - \delta \left( d \theta \right) \wedge d x \ \ ,$$ $$ \delta \left( \lambda d \theta \right) = \frac{1}{\lambda} \delta \left( d \theta \right) \ \ , \ \ d \theta \iota \delta \left( d \theta \right) = - \delta \left( d \theta \right) \ \ , \ \ d \theta \iota^p \delta \left( d \theta \right) = - p \iota^{p-1} \delta \left( d \theta \right) \ \ .
\end{equation}
The first property defines how $\delta \left( d \theta \right)$'s have to be used in order to perform form integration along the commuting directions $d \theta$'s; the second property reflects the usual property of the support of the Dirac distribution; the third and fourth properties implies that $\displaystyle \left| \delta \left( d \theta \right) \right| = 1 \mod 2$, i.e., $\delta \left( d \theta \right)$'s are odd objects and together with the fifth property they indicate that actually these are not really distributions, but rather \emph{de Rham currents}; the last two properties amounts for the usual integration by parts of the Dirac delta.

Given these properties it is possible to find a \virgolette top form" among integral forms as
\begin{equation}\label{ABRIFF}
	\omega_{top}^{(n|m)} = \omega \left( x , \theta \right) dx^1 \wedge \ldots \wedge dx^n \wedge \delta \left( d \theta^1 \right) \wedge \ldots \wedge \delta \left( d \theta^m \right) \ ,
\end{equation}
where $\omega \left( x , \theta \right)$ is a superfield. The space of $(n|m)$ forms is often indicated as $\mathpzc{B}er \left( \mathcal{SM} \right) \equiv \Omega^{(n|m)} \left( \mathcal{SM} \right)$ and is called the \emph{Berezinian bundle} (in analogy to the Determinant bundle of bosonic manifolds), since  the generator $\displaystyle \mathpzc{B}er^{(n|m)} = dx^1 \wedge \ldots \wedge dx^n \wedge \delta \left( d \theta^1 \right) \wedge \ldots \wedge \delta \left( d \theta^m \right)$ transforms with the Berezinian (i.e., the superdeterminant) of the Jacobian under any change of coordinates.

One can also consider a third class of forms, with non-maximal and non-zero number of deltas: \emph{pseudoforms}. A general pseudoform with $q$ deltas is locally given by
\begin{equation}\label{ABRIFG}
	\omega^{(p|q)} = \omega_{[a_1 \ldots a_r](\alpha_1 \ldots \alpha_s)[\beta_1 \ldots \beta_q]} \left( x , \theta \right) dx^{a_1} \wedge \ldots \wedge dx^{a_r} \wedge d \theta^{\alpha_1} \wedge \ldots \wedge d \theta^{\alpha_s} \wedge \delta^{(t_1)} \left( d \theta^{\beta_1} \right) \wedge \ldots \wedge \delta^{(t_q)} \left( d \theta^{\beta_q} \right) \ ,
\end{equation}
where we used the compact notation $\delta^{(i)} \left( d \theta \right) \equiv \left( \iota \right)^i \delta \left( d \theta \right)$. The form number is obtained as 
\begin{equation}\label{ABRIFH}
	p = r + s - \sum_{i=1}^q t_i \ ,
\end{equation}
since the contractions carry negative form number. The two quantum numbers $p$ and $q$ in eq. \eqref{ABRIFH} correspond to the {\it form} number and the {\it picture} number, respectively, and they range as $-\infty < p < +\infty$ and $0 \leq q \leq m$, so the picture number counts the number of delta's. If $q=0$ we have superforms, if $q=m$ we have integral forms, if $0<q<m$ we have pseudoforms.

As in conventional geometry, we can define the integral of a top form on a supermanifold\footnote{Actually, the form integration is not really on the supermanifold, but rather on the cotangent space $T^*\mathcal{SM}$. In order to avoid detailed discussions we will not linger on these themes.} as
 \begin{eqnarray}\label{ABRIFI}
I[\omega] = \int_{{\cal SM}} \omega_{top}^{(n|m)} = \int_{\cal SM} \omega_{top}^{(n|m)} [dx d\theta d(dx) d(d\theta)] \ ,
\end{eqnarray}
where the order of the integration variables is kept fixed and the \virgolette measure" $[dx d\theta d(dx) d(d\theta)]$ is invariant under coordinate transformations. We refer the reader to \cite{Witten:2012bg} for a complete discussion on the symbol $[dx d\theta d(dx) d(d\theta)]$. Here we simply recall that while $dx$ is ordinary Lebesgue integral, the integrations $d\theta$ and $d(dx)$ are Berezin integrals and the integration $d(d\theta)$ is algebraic, as discussed above. Performing the Berezin $d[dx]$ integrations and the algebraic $d[d \theta]$ ones in \eqref{ABRIFI}, it is then easy to check that $I[\omega]$ is nothing but the ordinary superspace integral  
\begin{eqnarray}\label{ABRIFJ}
I[\omega] = \int_{{\cal SM}} \omega(x, \theta) \left[ dx^1 \dots dx^n d\theta^1 \dots d\theta^m \right]
\end{eqnarray}
of the $\omega(x, \theta)$ superfield. In the present formulation the Stokes theorem for integral forms is also valid. 

By changing the $(1|0)$-forms $dx^a, d\theta^\a$ as 
\begin{equation}\label{ABRIFK}
dx^a \rightarrow E^a = E^a_m dx^m + E^a_\mu d\theta^\mu \qquad , \qquad d\theta^\a \rightarrow E^\a = E^\a_m dx^m + E^\a_\mu d\theta^\mu \ ,
\end{equation}
a top form $\omega^{(n|m)}$  transforms as
\begin{eqnarray}\label{ABRIFL}
\omega^{(n|m)} \rightarrow {\rm Ber}( E)  \, \omega(x,\theta) \, dx^1 \dots dx^n \, \delta(d\theta^1) \dots \delta(d\theta^m) \ ,
\end{eqnarray}
where $ {\rm Ber}(E)$ is the Berezinian of the supervielbeine $(E^a, E^\a)$. 

\subsection{Picture Changing Operators}

The strategy that we use for constructing integral forms and the corresponding supermanifold integrals is the following. Given a $(p|0)$-superform $\omega^{(p|0)}$ on a supermanifold ${\cal SM}$ of dimensions $(n|m)$ ($n \geq p$), its integration over a $p$-dimensional submanifold ${\cal N} \subset  {\cal SM}$ can be defined  as the integration on the entire supermanifold of the integral form $\omega^{(p|0)} \wedge \mathbb{Y}^{(n-p|m)}_{\cal N}$, where $\mathbb{Y}^{(n-p|m)}_{\cal N}$ is the \emph{Poincar\'e dual} to the immersion of ${\cal N}$ into ${\cal SM}$ \cite{Belo,Castellani:2015paa}. Precisely, if we denote $\omega^{(p|0)*} \equiv i^* \omega^{(p|0)}$ where $\displaystyle i : \mathcal{N} \hookrightarrow \mathcal{SM} $ is the  immersion of ${\cal N}$ into ${\cal SM}$ \footnote{Precisely, we consider ${\cal N} \subset {\cal M}$ where ${\cal M}$ is the bosonic component of  $ {\cal SM}$ known in the literature as the body.}, we define
\begin{eqnarray}\label{APCOA}
	\int_{\cal N} \omega^{(p|0)*} = \int_{\cal SM} \omega^{(p|0)} \wedge \mathbb{Y}^{(n-p|m)}_{\cal N} \ .
\end{eqnarray}
The second expression is the integral over the whole supermanifold of a $(n|m)$-dimensional top form to which we can then apply the usual Cartan calculus rules.
Operator $\mathbb{Y}^{(n-p|m)}_{\cal N}$ is also known as {\it Picture Changing Operator} (PCO), being related to a similar concept in string theory. 

The PCO in \eqref{APCOA} is independent of the coordinates, it only depends on the immersion through its homology class. Its main properties are:
\begin{equation}\label{APCOB}
	d{\mathbb Y}^{(n-p|m)}_{\cal N} = 0 \ \ , \ \ \mathbb{Y}^{(n-p|m)}_{\cal N} \neq d \Sigma^{(n-p-1|m)} \ ,
\end{equation}
and by changing the immersion $i$ to an homologically equivalent surface ${\cal N}'$, the new Poincar\'e dual $\mathbb{Y}^{(n-p|m)}_{\cal N'}$  differs from the original one by $d$-exact terms:
\begin{equation}\label{APCOC}
	\delta {\mathbb Y}^{(n-p|m)}_{\cal N} = {\mathbb Y}^{(n-p|m)}_{\cal N'} - {\mathbb Y}^{(n-p|m)}_{\cal N} = d \Lambda^{(n-p-1|m)} \ .
\end{equation} 
As a consequence of this, if $\omega^{(p|0)}$ is a closed form, then \eqref{APCOA} is automatically invariant under any change of the embedding (modulo boundary contributions). 
 
A notable example of application of this formalism is represented by the action of a rigid supersymmetric model: given a superform Lagrangian $\mathcal{L}^{(n|0)}$, we can lift it to an integral form by means of a PCO $\mathbb{Y}^{(0|m)}$ so that we can write the corresponding action as an integral on the whole supermanifold as
\begin{eqnarray}\label{APCOD}
	S = \int_{\cal SM} {\cal L}^{(n|0)} \wedge {\mathbb Y}^{(0|m)} \ .
\end{eqnarray} 
The $(n|0)$-form Lagrangian ${\cal L}^{(n|0)}$ is built using the rheonomic rules (see \cite{cube}) and turns out to be a function of dynamical superfields, the PCO ${\mathbb Y}^{(0|m)}$ instead contains only geometric data, for instance supervielbeine or coordinates themselves. If  $d  {\cal L}^{(n|0)} = 0$, we can change the PCO by exact terms without changing the action. This can be conveniently exploited for choosing for instance a PCO that possesses manifest symmetries.

This example has a natural generalization to supergravity. After the change \eqref{ABRIFK}, we can promote the flat forms generators (or flat supervielbeine) to dynamical fields and the action becomes 
\begin{eqnarray}\label{APCOE}
	S_{sugra} = \int_{{\cal SM}^{(n|m)}} {\cal L}^{(n|0)}(\Phi, E) \wedge  {\mathbb Y}^{(0|m)}(E) \ ,
\end{eqnarray}
where we have emphasised the dependence on the dynamical supervielbeine.

\subsection{The 2-dimensional Case.}

If we restrict to a $(2|2)$-dimensional supermanifold $\mathcal{SM}^{(2|2)}$, the \virgolette standard" PCO that is used to lift a $(2|0)$ superform to a top form is constructed by considering the projection on the base manifold, given by the constraints
\begin{equation}\label{AT2DCA}
	\theta = 0 = \bar{\theta} \ \ , \ \ d \theta = 0 = d \bar{\theta} \ \ .
\end{equation}
This lead to the component PCO
\begin{equation}\label{AT2DCB}
	\mathbb{Y}^{(0|2)} = \theta \delta \left( d \theta \right) \wedge \bar{\theta} \delta \left( \bar{d \theta} \right) \ .
\end{equation}
\eqref{AT2DCB} is trivially closed by using the properties listed in \eqref{ABRIFE} and it is not difficult to see that it is not exact. Once multiplied to a $(p|0)$-superform $\omega^{(p|0)}$ ($p\leq 2$, otherwise one would get automatically 0), the form is projected on its purely bosonic part:
\begin{equation}\label{AT2DCC}
	\omega^{(p|0)} \left( z , \bar{z} , \theta , \bar{\theta} , dz , d \bar{z} , d \theta , d \bar{\theta} \right) \wedge \mathbb{Y}^{(0|2)} = \omega^{(p|0)} \left( z , \bar{z} , 0 , 0 , dz , d \bar{z} , 0 , 0 \right) \wedge \mathbb{Y}^{(0|2)} \ .
\end{equation}
By using the supervielbeine defined in \eqref{oneA}, one can construct a different PCO, which is invariant under supersymmetry transformations as (we drop the \virgolette $\wedge$" symbols)
\begin{equation}\label{AT2DCD}
	\mathbb{Y}^{(0|2)} = V \iota \delta \left( \psi \right) \bar{V} \bar{\iota} \delta \left( \bar{\psi} \right) \ .
\end{equation}
We can directly see that this PCO differs from the component one by exact terms, by using \eqref{oneA} and \eqref{ABRIFE}:
\begin{equation}
	V \iota \delta \left( \psi \right) \bar{V} \bar{\iota} \delta \left( \bar{\psi} \right) = \left( dz + \theta d\theta \right) \iota \delta \left( d \theta \right) \left( d \bar{z} + \bar{\theta} d \bar{\theta} \right) \bar{\iota} \delta \left( d \bar{\theta} \right) = $$ $$ = - d z d \bar{z} \iota \delta \left( d \theta \right) \bar{\iota} \delta \left( d \bar{\theta} \right) - dz \iota \delta \left( d \theta \right) \bar{\theta} \delta \left( d \bar{\theta} \right) - \theta \delta \left( d \theta \right) d \bar{z} \bar{\iota} \delta \left( d \bar{\theta} \right) + \theta \delta \left( d \theta \right) \bar{\theta} \delta \left( d \bar{\theta} \right) = $$ $$ = \theta \delta \left( d \theta \right) \bar{\theta} \delta \left( d \bar{\theta} \right) + d \left[ \frac{\theta}{2} d z d \bar{z} \iota^2 \delta \left( d \theta \right) \bar{\iota} \delta \left( d \bar{\theta} \right)+ \frac{\theta}{2} dz \iota^2 \delta \left( d \theta \right) \bar{\theta} \delta \left( d \bar{\theta} \right) + \frac{\bar{\theta}}{2} \theta \delta \left( d \theta \right) d \bar{z} \bar{\iota}^2 \delta \left( d \bar{\theta} \right) \right] \ .
\end{equation}
With the same techniques one can directly verify that also the half-supersymmetric PCO constructed in \eqref{oneS} is cohomologically equivalent to the component one.

\end{document}